\documentclass{aa}     
\usepackage{graphics}
 
\def\cm{\,{\rm cm}}
\def\cc{\,{\rm cm^{-3}}}
\def\cm2{\,{\rm cm^{-2}}}

\def\kms{\,{\rm {km\,s^{-1}}}}
\def\kkms{\,{\rm {K\,km s^{-1}}}}

\def\C34S{\,{\rm C^{34}S}}
\def\co{\,{\rm CO}}
\def\13co{\,{\rm ^{13}CO}}
\def\h2{\,{\rm H_{2}}}

\def\sun{\odot}

\def\aua{{\rm A\&A} }
\def\auas{{\rm A\&AS} }
\def\apj{{\rm ApJ} }
\def\aj{{\rm AJ} }
\def\apjs{{\rm ApJS} }
\def\apjl{{\rm ApJL} }

\def\mnras{{\rm MNRAS} }

\begin{document}
 
   \thesaurus{03         
              (11.09.1 NGC 7331; 
               11.09.4; 
               11.19.2;
               11.19.6;
               13.19.1, 
               09.13.2)
             }
   \title{Molecular gas in the bulge and ring of NGC~7331}
 
   \subtitle{}
 
   \author{F.P. Israel
          \inst{1}
           and F. Baas
          \inst{1,2}
           }
 
   \offprints{F.P. Israel}
 
  \institute{Sterrewacht Leiden, P.O. Box 9513, NL 2300 RA Leiden,
             The Netherlands
  \and       Joint Astronomy Centre, 660 N. A'ohoku Pl., Hilo,
             Hawaii, 96720, USA}

   \date{Received 6 January 1999; accepted 5 August 1999}
 
   \maketitle
 
   \begin{abstract}

CO emission from the Sb(rs)I-II galaxy NGC 7331 has been mapped in the 
$J$=2--1 transition with a 21$\arcsec$ beam over an area 3.5$\arcmin$ by 
1.3$\arcmin$. A relatively low contrast enhancement of molecular line 
emission occurs in a ringlike zone at a distance of approximately 3.5 kpc 
from the center; there is no evidence for a pronounced central hole. The 
ring is located at the edge of the region of rigid rotation and roughly 
coincides with an inhomogeneous ring of nonthermal radio continuum 
emission. It is well inside the radius of maximum rotational velocity.

The intensities of the 492 GHz [CI] line and various $^{12}$CO and $^{13}$CO 
transitions observed towards the center and two outlying positions are
modelled by multiple molecular gas components: low-density gas at a
kinetic temperature $T_{\rm kin} \approx 10$ K, and high-density gas
at both $T_{\rm kin} \approx 10$ K and $T_{\rm kin} \approx 20$ K. The
molecular gas must be distributed in clumpy or filamentary form. 
The CO-to-H$_{2}$ conversion factor $X$ applicable to the bulge is only
half that applicable to the ring and beyond. The latter is still
significantly lower than $X_{\rm Milky Way}$. Molecular hydrogen is
the dominant mass contributor to the interstellar medium in the bulge
and in the ring. Far-infrared emission from dust peaks inside the ring
at 100$\mu$m (warm dust), and in the ring at 850$\mu$m (colder dust). 
Beyond the ring, neutral atomic hydrogen is dominant.
Inferred total hydrogen mass densities in the ring are about twice those 
in the bulge. Interstellar gas to dynamical mass ratios are of order 
1$\%$ in the bulge, about 1.5$\%$ in the ring followed by a rise to 3$\%$. 
The bulge gas may have originated in mass loss from bulge stars; 
in that case, the molecular ring is probably caused by a decrease in 
evacuation efficiency at the bulge outer edge.


\end{abstract}
 
\section{Introduction}

NGC~7331 is an isolated spiral galaxy of type Sb(rs)I-II with prominent 
dust lanes close to its centre (Kormendy $\&$ Norman 1979; Sandage $\&$ 
Tammann 1987). Table 1 summarizes the relevant parameters of NGC~7331. 
A ringlike distribution of dust surrounding the bulge was suggested by 
Telesco et al. (1982). Such a distribution is also apparent in radio 
continuum maps (Cowan et al. 1994), but only vaguely 
in HI maps (Bosma 1978; Begeman 1987). A strong ring signature in CO 
emission was claimed by Young $\&$ Scoville (1982). 

\begin{table}
\caption[]{NGC~7331 parameters}
\begin{flushleft}
\begin{tabular}{ll}
\hline
Type$^{a}$     	 			& Sb(rs)I-II \\
Optical Centre:				& \\
R.A. (1950)$^{b}$ 	 		& 22$^{h}$34$^{m}$47.7$^{s}$  \\
Decl.(1950)$^{b}$        		& 34$^{\circ}$09$\arcmin$35$\arcsec$ \\
Radio Centre :				& \\
R.A. (1950)$^{c}$ 	 		& 22$^{h}$34$^{m}$46.6$^{s}$  \\
Decl.(1950)$^{c}$        		& 34$^{\circ}$09$\arcmin$21$\arcsec$ \\
$V_{\rm LSR}^{d}$    			& 831 $\kms$ \\
Distance $D^{e}$          		& 14.3 Mpc \\
Inclination $i^{f}$ 			& 74.8$^{\circ}$ \\
Position angle $P^{f}$     		& 167$^{\circ}$ \\
Luminosity $L_{\rm B}^{g}$  		& $5.0 \times 10^{10}$ L$_{\rm B\odot}$ \\
Scale           			& 14.5 $\arcsec$/kpc \\
\hline
\end{tabular}
\end{flushleft}
Notes to Table 1:\\
$^{a}$ RSA (Sandage $\&$ Tammann 1987) \\
$^{b}$ Dressel $\&$ Condon (1976) \\
$^{c}$ Begeman (1987); Cowan et al. (1994) \\
$^{d}$ Corresponding to V$_{\rm Hel}$ = 820 $\kms$ (Begeman 1987) \\
$^{e}$ Tully (1988); corresponds to H$_{\rm o}$ = 75 $\kms$ \\
$^{f}$ HI-derived parameters from Begeman (1987) \\
$^{g}$ Begeman (1987) rescaled to D = 14.3 Mpc \\
\end{table}

Several galaxies are thought to host a molecular ring structure (see 
e.g. Braine et al. 1993), such as the ring in our own Galaxy 
discovered by Scoville $\&$ Solomon (1975) and the one in M~31 
(Stark 1979; Dame et al. 1994; Koper 1993). The latter serves to 
illustrate a dilemma commonly facing the interpretation of CO maps,
especially those of highly inclined galaxies where rings are most easily 
discerned: the conspicuous molecular structure may in fact consist of 
spiral arm segments that only in projection suggest a ringlike structure.
Two-dimensional CO mapping of NGC~7331 by von Linden et al. 1996) and 
Tosaki $\&$ Shioya (1997) support the latter interpretation. Although 
molecular rings have been identified in or claimed for other galaxies, 
the case of NGC~7331 is of interest because it resembles M~31 in being a 
large spiral galaxy of relatively early type, containing a prominent 
stellar bulge. It has also been claimed to have, as M~31, very little 
CO emission inside its molecular ring (Young $\&$ Scoville 1982; Tosaki 
$\&$ Shioya, 1997). NGC~7331 even resembles M~31 in its high inclination 
(75$^{\circ}$ and 77$^{\circ}$ respectively -- Arp $\&$ Kormendy 1972, 
Sandage $\&$ Tammann 1987). Its radio structure is a stronger version of 
that of M~31 (Cowan et al. 1994). NGC~7331 also contains a clear, but 
patchy radio continuum ring. Inside the ring, little or no radio 
emission is found, except for a compact nuclear source. The luminosity of 
this source is 3--4 times that of Sgr A, and a thousand times stronger than 
the nucleus of M~31. It is associated with a nuclear X-ray source 
(Stockdale et al. 1998). Ringlike distributions of interstellar dust are also 
seen at mid-infrared (Smith 1998) and submillimeter (Bianchi et al. 1998) 
wavelengths, but they are not nearly as evident at the far-infrared 
wavelengths inbetween (Smith $\&$ Harvey 1996; Alton et al. 1998).

An unusual characteristic of NGC~7331 is the rather low $J$=2--1/$J$=1--0 
CO transitional ratio of 0.5--0.7 reported by Braine et al. (1993) and 
von Linden et al. (1996). This is quite different from most other 
galaxies observed in CO, where the two transitions are usually of 
similar strength. However, such low transitional ratios have also been 
found for individual dark clouds in the central parts of M~31 (Allen $\&$
Lequeux 1993). These were interpreted as evidence for very cold ($T_{\rm kin} 
<$ 5 K) and tenuous ($n \approx$ 100 cm$^{-3}$) molecular clouds by
Loinard et al. (1995), but Israel et al. (1998) showed that they are
more likely caused by filamentary gas at temperatures $T_{\rm kin} \approx
10$ K present at both low and high densities. Given the similarities between
NGC~7331 and M~31, it is of interest to investigate whether such a state of 
affairs also applies to the central region of NGC~7331.

In this paper, we present a fully sampled map of NGC 7331 in the $J$=2--1 
$^{12}$CO transition over an area of 1.3$\arcmin$ by 3.5$\arcmin$.
In addition, we have measured the first three $^{12}$CO and $^{13}$CO 
transitions as well as the 492 GHz CI transition towards the central region
of the galaxy, allowing us to narrow down the permitted range of the 
apparently unusual physical conditions in the center.

\section{Observations}

Details relevant to the observations are listed in Table 2; the system
temperatures given are the means for the respective runs. Observations in 
the $J$=1--0 transition were obtained with the IRAM 30 m telescope in 
service mode, at the optical and radio centre positions respectively, 
separated by $\Delta \alpha$ = 13.7$\arcsec$, $\Delta \delta$ = 
14.0$\arcsec$. The $^{13}$CO observations were bracketed by the $^{12}$CO 
observations.

\begin{table}
\caption[]{Observations Log}
\begin{flushleft}
\begin{tabular}{lccccc}
\hline
\noalign{\smallskip}
Transition & Date    	& Freq	& $T_{\rm sys}$ & Beamsize    	& $\eta _{\rm mb}$ \\
	   & (MM/YY) 	& (GHz)	& (K)		  & ($\arcsec$) & 	       \\
\noalign{\smallskip}
\hline
\noalign{\smallskip}
\multicolumn{6}{l}{$^{12}$CO} \\
\noalign{\smallskip}
$J$=1--0   & 07/97	& 115	& 350	  	& 21	        &  0.74 \\
$J$=2--1   & 11/91   	& 230	& 850	  	& 21	        &  0.63 \\
	   & 08-10/92 	& 	& 600	  	& 21		&  0.63 \\
	   & 01/96   	& 	& 500	  	& 21	        &  0.69 \\
	   & 12/97  	&	& 260     	& 21		&  0.69 \\
$J$=3--2   & 05/93   	& 345	& 720	  	& 14	        &  0.53 \\
	   & 07/95   	& 	& 900     	& 14		&  0.58 \\
\noalign{\smallskip}
\hline
\noalign{\smallskip}
\multicolumn{6}{l}{$^{13}$CO} \\
\noalign{\smallskip}
$J$=1--0   & 07/97	& 110	& 200	  	& 21 		&  0.74 \\
$J$=2--1   & 09/93	& 220	& 660	  	& 21	        &  0.63 \\
	   & 12/97	&	& 320     	& 21		&  0.69 \\
\noalign{\smallskip}
\hline
\noalign{\smallskip}
\multicolumn{6}{l}{CI} \\ 
$^{3}$P$_{1}$--$^{3}$P$_{0}$ & 11/96 & 492	& 1900	  	& 10 &  0.53 \\
\noalign{\smallskip}
\hline
\end{tabular}
\end{flushleft}
\end{table}

\begin{table}
\caption[]{CO and CI line intensities in NGC~7331}
\begin{flushleft}
\begin{tabular}{llcrc}
\hline
\noalign{\smallskip}
\multicolumn{2}{l}{Transition}	& Resolution  & $T_{\rm mb}$ & $\int T_{\rm mb}$d$V$ \\
& 				& ($\arcsec$) & (mK)	     & (K $\kms$) \\
\noalign{\smallskip}
\hline
\noalign{\smallskip}
\multicolumn{5}{c}{A : $\alpha$ = 22:34:46.6; $\delta$ = 34:09:21} \\
\noalign{\smallskip}
\hline
\noalign{\smallskip}
$J$=1--0 & $^{12}$CO   & 21	      & 162	      & 26$\pm$2  \\
\noalign{\smallskip}
	 & $^{13}$CO   & 21	      &  21	      & 3.9$\pm$0.4 \\
\noalign{\smallskip}
$J$=2--1 & $^{12}$CO   & 21	      & 100	      & 14$\pm$2 \\
\noalign{\smallskip}
	 & $^{13}$CO   & 21	      &  13	      & 2.5$\pm$0.6 \\
\noalign{\smallskip}
$J$=3--2 & $^{12}$CO   & 14           &  43	      & 7.5$\pm$1.3 \\
	 &	       & 21	      &  49	      & 8.8$\pm$1.5 \\
\noalign{\smallskip}
$^{3}$P$_{1}$--$^{3}$P$_{0}$ & CI & 10	&  30	      & 1.9$\pm$0.3 \\
\noalign{\smallskip}
\hline
\noalign{\smallskip}
\multicolumn{5}{c}{B : $\alpha$ = 22:34:47.7; $\delta$ = 34:09:35} \\
\noalign{\smallskip}
\hline
\noalign{\smallskip}
$J$=1--0 & $^{12}$CO   & 45$^{\rm a}$ &  47	      &  7$\pm$1  \\
	 &	       & 33$^{\rm b}$ & 170	      & 27$\pm$4  \\
	 &	       & 21	      & 245	      & 25$\pm$2  \\
\noalign{\smallskip}
	 & $^{13}$CO   & 21	      &  27	      & 3.3$\pm$0.4 \\
\noalign{\smallskip}
$J$=2--1 & $^{12}$CO   & 21	      & 109	      & 12$\pm$2  \\
\noalign{\smallskip}
	 & $^{13}$CO   & 21	      &  13	      & 1.8$\pm$0.3 \\
\noalign{\smallskip}
$J$=3--2 & $^{12}$CO   & 14           & 100	      & 10$\pm$2 \\
	 &	       & 21$^{\rm c}$ &  --	      & 10$\pm$4   \\
\noalign{\smallskip}
\hline
\noalign{\smallskip}
\multicolumn{5}{c}{C : $\alpha$ = 22:34:48.9; $\delta$ = 34:09:34} \\
\noalign{\smallskip}
\hline
\noalign{\smallskip}
$J$=1--0 & $^{12}$CO   & 21$^{\rm d}$ &  --	      & 10$\pm$2     \\
\noalign{\smallskip}
$J$=2--1 & $^{12}$CO   & 21	      &  36	      & 3.5$\pm$1  \\
\noalign{\smallskip}
	& $^{13}$CO    & 21	      &   5:	      & 0.6$\pm$0.3  \\
\noalign{\smallskip}
$J$=3--2 & $^{12}$CO   & 14           &  53	      & 2.5$\pm$0.5  \\
	 &	       & 21$^{\rm c}$ &  --	      & 3.5$\pm$1.5 \\
\noalign{\smallskip}
\hline
\end{tabular}
\end{flushleft}
Notes to Table 3: \\ 
a. From Young et al. (1995); b. From Elfhag et al. (1996); \\
c. Extrapolated value; d. From von Linden et al. (1996)
\end{table}

\begin{figure*}

\vspace{-2.5cm}
\hspace{-0.6cm}
\resizebox{12cm}{!}{\rotatebox{270}{\includegraphics*{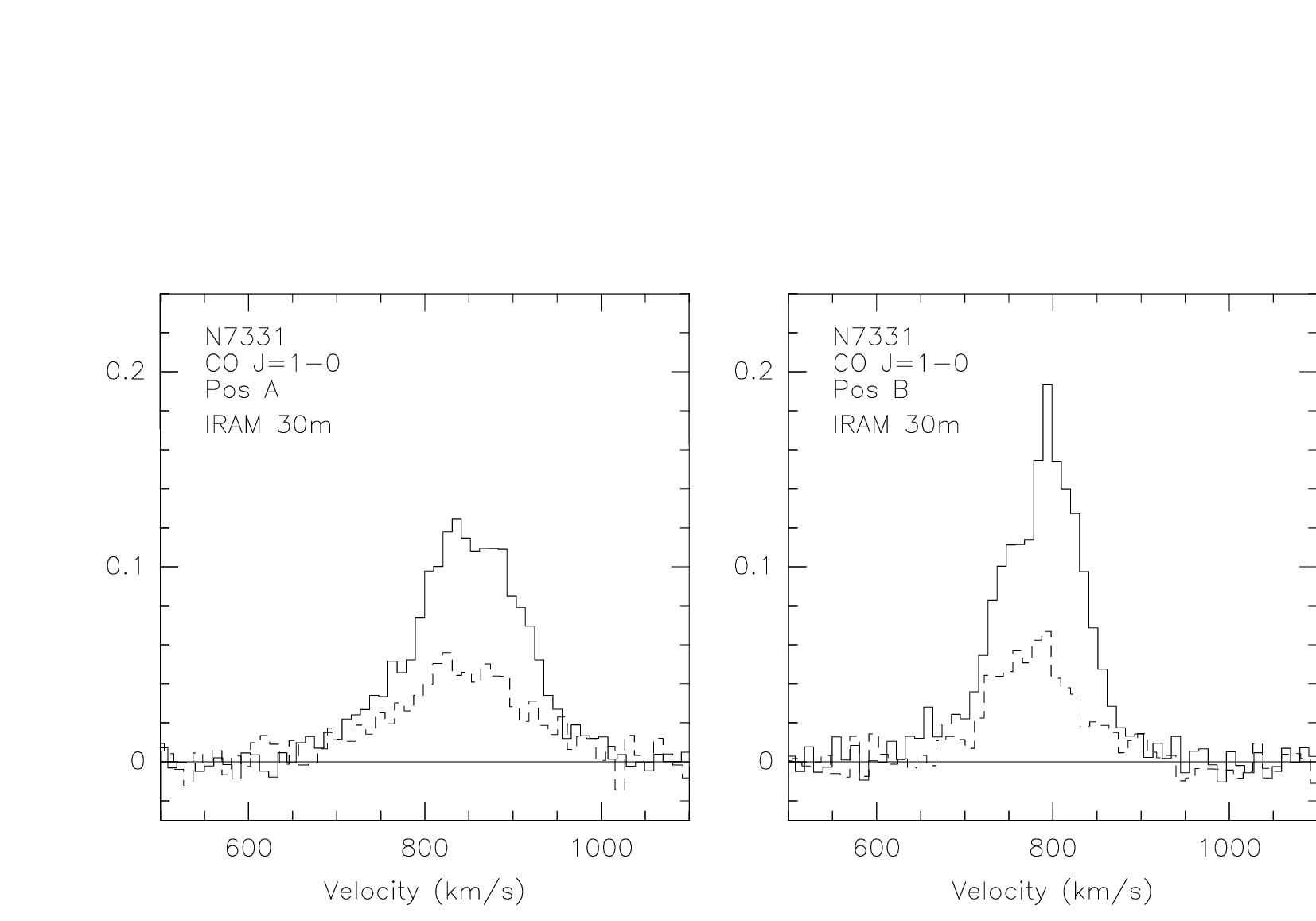}}}

\vspace{-6.2cm}
\resizebox{18cm}{!}{\rotatebox{270}{\includegraphics*{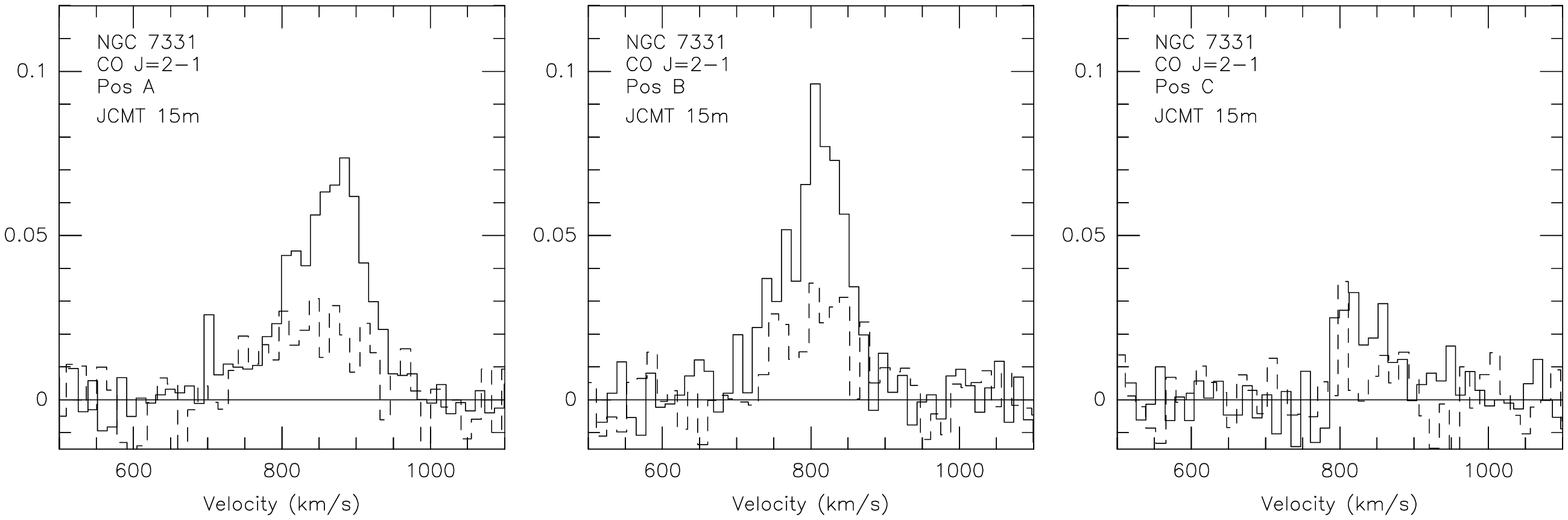}}}

\vspace{-0.28cm}
\hspace{0.25cm}
\resizebox{5.68cm}{!}{\rotatebox{270}{\includegraphics*{N7331F1c.ps}}}
\hspace{-0.15cm}
\resizebox{5.68cm}{!}{\rotatebox{270}{\includegraphics*{N7331F1d.ps}}}
\hspace{-0.15cm}
\resizebox{5.68cm}{!}{\rotatebox{270}{\includegraphics*{N7331F1e.ps}}}

\vspace{0.4cm}
\hspace{0.24cm}
\resizebox{5.68cm}{!}{\rotatebox{270}{\includegraphics*{N7331F1f.ps}}}
\hspace{1.0cm}\parbox[t]{10cm}{\caption[]
{Spectra observed towards positions A, B and C in NGC~7331 (see Sect. 2).
Top row: $J$=1--0 CO; second row: $J$=2--1 CO; third row: $J$=3--2 CO;
bottom row: [CI]. 
Horizontal scale is LSR velocity in $\kms$, vertical scale is 
$T_{A}^{*}$ in K. To convert to $T_{\rm mb}$, multiply $T_{A}^{*}$ by 
1.35 ($J$=1--0 CO), 1.45 ($J$=2--1 CO), 1.72 ($J$=3--2 CO) and 1.89 
([CI]), respectively. \\
Note: $\co$ is marked by solid lines, $\13co$ multiplied by three
is marked by dashed lines.}} 
\end{figure*}

All other observations were carried out with the 15m James Clerk Maxwell 
Telescope (JCMT) on Mauna Kea (Hawaii) \footnote{The James Clerk Maxwell 
Telescope is operated on a joint basis between the United Kingdom Particle 
Physics and Astrophysics Council (PPARC), the Netherlands Organisation for 
Scientific Research (NWO) and the National Research Council of Canada (NRC).}.
Up to 1993, we used a 2048 channel AOS backend 
covering a band of 500 MHz ($650\kms$ at 230 GHz). After that year, the 
DAS digital autocorrelator system was used in bands of 500 and 750 MHz. 
Resulting spectra were binned to resolutions of 4 -- 10 $\kms$, except
for the 492 GHz CI spectrum which was binned to 18 $\kms$ on order to obtain
a sufficiently high signal-to-noise ratio. Only linear baseline corrections
were applied to the spectra. All spectra were scaled to a main-beam 
brightness temperature, $T_{\rm mb}$ = $T_{\rm A}^{*}$/$\eta _{\rm mb}$;
relevant values for $\eta _{\rm mb}$ are given in Table 2. Spectra are 
shown in Fig. 1 and summarized in Table 3. Position A (indicated in 
Figure 2) is that of the radio nucleus, the actual galaxy center. 
Along the minor axis, the beam just fills the space between the ring
components. Positions B and C are close to the minor axis at deprojected 
radii $R$ = 4.5 kpc and $R$ = 8.2 kpc respectively; position B is centered
on the CO ring. Because of the pronounced tilt of the
galaxy, 21$''$ circular observing beams sample elongated ellipses in the
plane of the galaxy, covering a range of 5.55 kpc in the minor axis
direction. The beams covering positions B and C overlap (see Figure 2);
positions A and B are essentially independent.

For the $J$=2--1 mapping observations, the integration time was typically 
400 seconds per spectrum, on and off the source. After binning to a
velocity resolution of 5 $\kms$, the resulting r.m.s. noise and 
baseline deviations were of the order of 20 mK over most of the band. 

\begin{figure}
\resizebox{11cm}{!}{\rotatebox{270}{\includegraphics*{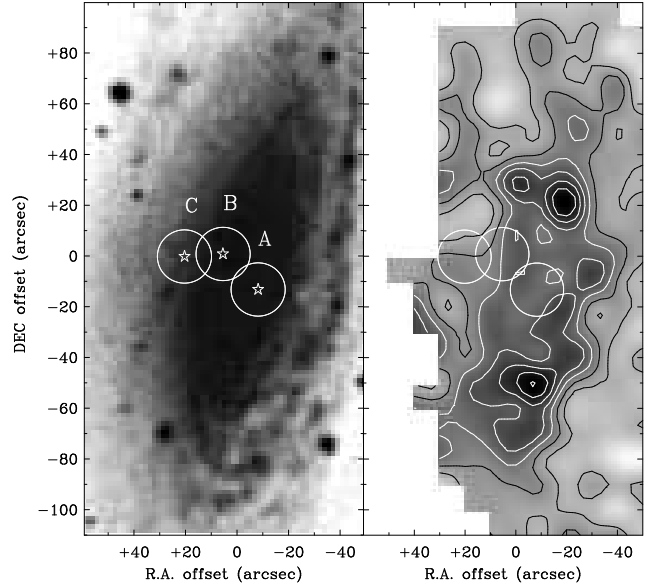}}}
\caption[]{Left: Digitized sky survey image of NGC~7331 with observed
positions A, B and C (see Table 3) marked. Size of circle corresponds to 
230 GHz beamsize. Right: Distribution of $J$=2-1 $^{12}$CO emission in 
NGC~7331, integrated over a velocity range V$_{\rm LSR}$ = 530 - 1130 km 
s$^{-1}$. Contours are in steps of $\int T{\rm mb}$d$V$ = 4 K km s$^{-1}$. 
Observed positions A, B and C are indicated by circles. 
Table 1. }
\end{figure}

\section{Results}

\subsection{CO distribution}

In the $J$=1--0 and $J$=2--1 transitions, the integrated intensities of 
both CO isotopes are rather similar for positions A and B, indicating a 
fairly smooth central distribution of relatively strong CO. This is 
consistent with the large (50$\%$) fraction of flux found to be missing
by Tosaki $\&$ Shioya (1997) in their $J$=1--0 CO interferometer map. 
The results presented here and by von Linden et al. (1996) are, however, 
inconsistent with the folded major axis profile obtained by Young $\&$ 
Scoville (1982) and shown in more detail by Young et al. (1995), which 
exhibits both a pronounced lack of CO at the centre and a strong peak at 
a radial distance of 45$\arcsec$ (3 kpc). From the data in Table 2 and 
the results obtained by von Linden et al. (1996), it appears that 
the central integrated value given by Young et al. (1995) is too low by 
more than a factor of two. Thus, {\it there is no significant central 
`hole' in the distribution of CO emission}, at least not on the scale 
of our 20$''$ beam.

A full-resolution contour map of the $J$=2--1 CO intensity, integrated 
over the velocity range of 530 to 1130 $\kms$ is shown in Fig. 2. The map  
shows a good overall resemblance to the $J$=1--0 CO map obtained at slightly
lower resolution by von Linden et al. (1996). This is also true for the major 
axis position-velocity diagram (not shown here). In Fig. 2, the elliptical 
outline of a low-contrast ring around the center can be discerned; peaks of 
CO emission occur at positions 30$\arcsec$ north and 50$\arcsec$ south. The 
latter two more or less correspond to the radial distance of the molecular 
ring proposed by Young $\&$ Scoville (1982). The map covers the brightest 
part of the optical image also whown in Figure 2 (for better images, see
panel 40 in the atlas by Sandage $\&$ Bedke 1988). In this image, a large,
overexposed bulge is surrounded by 
dust lanes and irregular spiral arms traced by HII regions. The CO maxima 
at $\Delta\delta$ = -50$''$ and +30$''$ fall on either side of the bulge, 
and the ring traces dusty spiral arms close to the bulge, 
especially on the western side. Most of the reddening of NGC~7331 occurs 
in this western spiral arm (Telesco et al. 1982; Bianchi et al. 
1998). On the eastern side of the CO map, faint emission due to a more 
distant major spiral arm is seen as well. The HI map obtained by Begeman 
(1987) at a very similar resolution shows an incomplete `ring' of neutral 
hydrogen. The CO emission from the outlying spiral arm coincides with a 
relatively bright part of this HI `ring'. Most of the CO emission is, 
however, well inside it and coincides with the radio continuum ring mapped 
by Cowan et al. (1994). The main CO peaks are at the northern and southern 
extremities of the radio continuum ring. 

In the velocity-integrated single-dish CO map (Fig. 2), the ring is only
weakly visible. Its presence is more clearly revealed in the interferometer 
map of Tosaki $\&$ Shioya (1997) and in Fig. 3. This figure shows the 
distribution of $J$=2--1 $^{12}$CO over the same region, but now integrated 
over velocity bins of 40$\kms$ only. Between velocities V$_{\rm LSR}$ = 600 
and 1000 $\kms$, the maps show a double structure. The 
double-peak structure seen in Fig. 3 extends over most of the part of 
NGC~7331 characterized by rigid rotation (cf. von Linden et al. 1996).
A similar pattern, with the limitations imposed by interferometric techniques,
is also evident in the channel maps published by Tosaki $\&$ Shioya (1997).

\subsection{Line ratios}

The $^{12}$CO/$^{13}$CO isotopical ratios of 6 -- 7 are somewhat low 
compared to typical values around 10 found in most other external galaxies, 
but given the low IRAS $f_{60}/f_{100}$ ratio of 0.3 (Rice et al. 1988) this 
is in line with the results obtained by Aalto et al. (1991). We find 
a $^{12}$CO $J$=2--1/$J$=1--0 ratio of 0.54$\pm$0.10 in the center, 
consistent with the results obtained by Braine et al. (1993) and von Linden 
et al. (1996). Both $J$=1--0/$J$=2--1 intensities and ratio decrease away
from the center. The CO transitional ratios observed in NGC~7331 are rather 
different from those of most other galaxies, where the lower $J$ 
transitions usually have similar velocity-integrated intensities
(cf. Israel $\&$ van der Werf 1996). In contrast, CO intensities in the 
centre of NGC~7331 decrease rapidly with increasing $J$ level (Table 3). 
The $J$=2--1/$J$=1--0 ratios suggest subthermal excitation either at 
relatively low excitation temperatures or at very low column densities. 
However, the observed $J$=3--2/$J$=2--1 ratios of 0.6 or higher indicate
the presence of a certain amount of warm molecular gas. Very low 
temperatures are also unlikely because they imply $\co/\13co$ ratios 
substantially closer to unity than is observed. We also note, in Table 4, 
a similarity between the ratios applicable to the D~478 cloud in the central 
parts of M~31 and to the emission from NGC~7331, notwithstanding the factor
of 400 difference in beam surface area.

\begin{figure*}
\resizebox{16cm}{!}{\rotatebox{270}{\includegraphics*{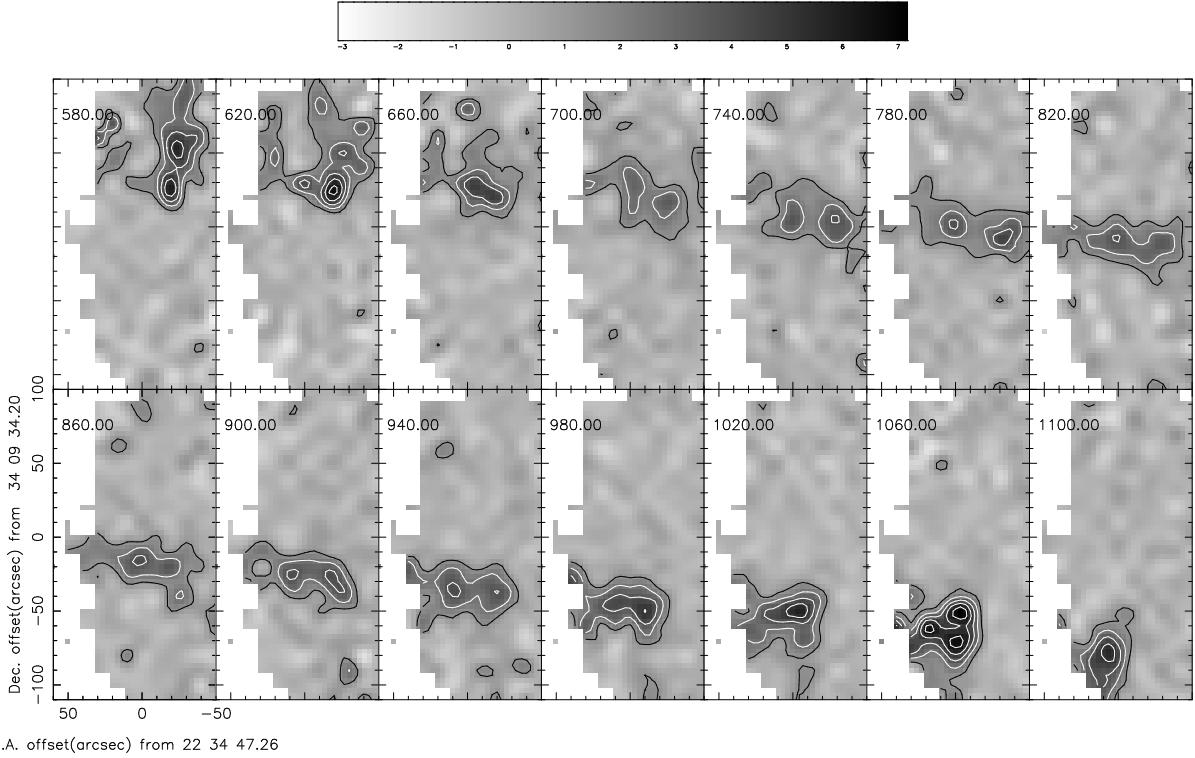}}}
\caption[]{$J$=2-1 $^{12}$CO channel maps of NGC~7331. Emission is 
integrated over velocity bins of 40 $\kms$; central velocities are
marked in the panels. Contours are in steps of $\int T_{\rm mb}$d$V$ =
2 K km s$^{-1}$.
}
\end{figure*}

\section{Analysis and discussion}

\subsection{Modelling of observed line intensities}

\begin{table*}
\caption[]{Integrated line ratios in the centre of NGC~7331}
\begin{flushleft}
\begin{tabular}{llcccc}
\hline
\noalign{\smallskip}
Transitions	&		& pos. A  	& pos B. 	& pos. C 	& D~478$^{a}$ \\
\noalign{\smallskip}
\hline
\noalign{\smallskip}
$^{12}$CO & (2--1)/(1--0)	& 0.54$\pm$0.10	& 0.51$\pm$0.10	& 0.35$\pm$0.11	& 0.42$\pm$0.08 \\
	  & (3--2)/(2--1)	& 0.63$\pm$0.15	& 0.78$\pm$0.35	& 1.0$\pm$0.41	& 0.33$\pm$0.10 \\
\noalign{\smallskip}
$^{13}$CO & (2--1)/(1--0) 	& 0.64$\pm$0.2	& 0.55$\pm$0.1	&   ---		& 0.45$\pm$0.15 \\
\noalign{\smallskip}
$^{12}$CO/$^{13}$CO & (1--0)	& 6.7$\pm$1.1 	& 7.8$\pm$1.2	&   ---	      	& 8.8$\pm$1.8 \\
		    & (2--1)	& 5.6$\pm$1.6 	& 7.2$\pm$1.6	& 5.8$\pm$3.3	& 8.4$\pm$1.7 \\
\noalign{\smallskip}
CI/CO(2--1)		& 	& 0.21$\pm$0.07 &    ---	&   ---		& 0.24$\pm$0.05 \\
\noalign{\smallskip}
\hline
\end{tabular}
\end{flushleft}
Note: a. Dark cloud in M~31; see Allen et al. (1995); Loinard $\&$ Allen
(1998); Israel et al. (1998).
\end{table*}

\begin{table*}
\caption[]{Model parameters for NGC~7331}
\begin{flushleft}
\begin{tabular}{lccccc}
\hline
\noalign{\smallskip} 
Model & \multicolumn{3}{c}{Cold Component} 	    	   	 & \multicolumn{2}{c}{Warm Component} \\
     & Kinetic	      & Gas	      & CO Column 	   	 & Kinetic 		& Gas 	      \\
     & Temperature    & Density       & Density   	   	 & Temperature 		& Density     \\
     & T$_{\rm kin}$  & n(H$_{2}$)    & N(CO)/dV	    	 & T$_{\rm kin}$ 	& n(H$_{2}$)  \\
     & (K)     	      & ($\cc$)       & ($10^{17}\cm2 (\kkms)^{-1}$) & (K) 		& ($\cc$)    \\
\noalign{\smallskip}
\hline
\noalign{\smallskip}
1    & 10/10$^{a}$    & 100/3000      & 0.7/7.0			 &  30			& 3000  \\
2    &  10	      & 300 	      &   1.0       		 &  40			& 1000  \\
3    &  10	      & 300 	      &   1.0     		 &  30			& 1000 	\\
4    &  10	      & 300	      &   1.0		  	 &  20			& 1000	\\
5    &  10	      & 100 	      &   1.0       		 &  30			& 1000  \\
6    &  10	      & 300 	      &   0.1     		 &  30			& 1000 	\\
7    &  10	      & 100  	      &   0.1   		 &  30			& 1000  \\
\noalign{\smallskip}
\hline
\end{tabular}
\end{flushleft}
Note: $^{a}$ Cold component assumed to consist of both low and high density gas
as in D~478; see Israel et al. (1998).
\end{table*}

\begin{table*}
\caption[]{CO and C in NGC~7331}
\begin{flushleft}
\begin{tabular}{lccccccccc}
\hline
\noalign{\smallskip} 
Model & Predicted    & \multicolumn{3}{c}{Beam-Averaged}    & Conversion & \multicolumn{2}{c}{Mass per Beam} & \multicolumn{2}{c}{Face-on } \\
      & CI Intensity & \multicolumn{3}{c}{Column Densities} & Factor $X$ & & & \multicolumn{2}{c}{Mass Density} \\
      & $\int T_{\rm mb}$d$V$ & $N(CO)$	& $N(C)$    & $N({\it \h2})$		  &  & $M(\h2)$  & $M_{\rm gas}$ & $\sigma(\h2)$ & $\sigma_{\rm gas}$ \\
      & ($\kkms$) & \multicolumn{3}{c}{($10^{20} \cm2)$} & ($10^{20} \cm2/\kkms$) & \multicolumn{2}{c}{($10^{7}$ M$_{\odot}$)} & \multicolumn{2}{c}{(M$_{\odot}$/pc$^{-2}$)} \\
\noalign{\smallskip}
\hline
\noalign{\smallskip}
& \multicolumn{5}{l}{Position A; $N_{H}/N_{C}$ = 1700; $N(HI) = 5 \times 10^{20} \cm2$} \\
1     &   2	     & 0.011	& 0.006	    & 12		     	  & 0.4        &  3	   &  5		   &  5		  &  8 \\
2     &   4	     & 0.012	& 0.040	    & 42		     	  & 1.6        & 11	   & 15		   & 17		  & 24 \\
3     &   4	     & 0.009	& 0.041	    & 40		     	  & 1.5        & 10	   & 14		   & 16		  & 23 \\
4     &   4	     & 0.008      & 0.042	    & 40			  & 1.5	       & 10	   & 14		   & 16		  & 23 \\
5     &   4	     & 0.009	& 0.051	    & 48		     	  & 1.9        & 13	   & 18		   & 20		  & 28 \\
6     &   8	     & 0.005	& 0.045	    & 39		     	  & 1.5        & 10	   & 15		   & 16		  & 23 \\
7     &   9	     & 0.008	& 0.089	    & 80		     	  & 3.1        & 21	   & 29	 	   & 33		  & 46 \\
\noalign{\smallskip}
\hline
\noalign{\smallskip}
& \multicolumn{5}{l}{Position B; $N_{H}/N_{C}$ = 2300; $N(HI) = 9 \times 10^{20} \cm2$} \\
1     &  2.5	     & 0.011	& 0.007	    & 16		     	  & 0.6        &  4	   &  7		   &  7		  & 12 \\
3     &   4	     & 0.006	& 0.030	    & 37		     	  & 1.5        & 10	   & 15		   & 15		  & 23 \\
\noalign{\smallskip}
\hline
\noalign{\smallskip}
& \multicolumn{5}{l}{Position C; $N_{H}/N_{C}$ = 3000; $N(HI) = 9 \times 10^{20} \cm2$} \\
1     &  0.7	     & 0.004	& 0.003	    &  8		     	  & 0.8        &  2	   &  4		   &  3		  &  7 \\
3     &  1.3	     & 0.002	& 0.013	    & 19		     	  & 1.9        &  5	   &  8		   &  8		  & 13 \\
\noalign{\smallskip}
\hline
\end{tabular}
\end{flushleft}
\end{table*}

We have modelled the observed intensities and their ratios by assuming the
presence of two molecular gas components of different temperature and
density, a relatively cold component dominating the $J$=1--0 emission and 
a warmer component becoming progressively more important in the higher 
transitions. We have used the radiative transfer models from the Leiden 
astrochemistry group (Jansen 1995; Jansen et al. 1994);
we included a background radiation field of $T_{\rm bg}$ = 2.73 K. 
In these models kinetic temperature, molecular hydrogen density and CO 
respectively C column densities function as input parameters. A further
constraint is provided by the chemical models discussed by Van Dishoeck
$\&$ Black (1988) which show a strong dependence of the $N(C)/N(CO)$
column density ratio around molecular hydrogen column densities of
about 10$^{21} \cm2$. Above $N(\h2) = 2 \times 10^{21}$, virtually all
carbon is in CO, whereas below $N(\h2) = 2 \times 10^{20} \cm2$ 
virtually all carbon is in C. The CI and CO intensities observed in 
position A, together with the sensitivity of the C/CO ratio to $\h2$ 
column density provide rather stringent constraints on the models acceptable 
for at least the center of NGC~7331.
 
Although the available data do not allow a precise and unique determination 
of the physical condition of the molecular gas in NGC~7331, they serve well 
to constrain the parameter space of possible solutions. For instance, cold 
component CO column densities cannot exceed $N(CO)$ $\approx 10^{17} \cm2$ and 
warm component kinetic temperatures cannot be lower than 20 K without
conflicting with observed $^{12}$CO and $^{13}$CO ratios. In Table 5
we list a few representative models. Note that the CO column densities 
listed in Table 5 are those of a single model cloud. We assume a homogeneous
population of such model clouds, non-shadowing in position-velocity
space, so that the actual galaxy beam-averaged column density is the 
sum of the model cloud column densities in the beam multiplied by their 
beam filling factor. 

In models 4 through 7 we vary the
cold component input parameters, and assume a warm component of 30 K
and density $n(\h2)$ = 1000 $\cc$. In models 2 and 3, we have changed the
warm component temperature to 40 K and 20 K, and in model 1  we assume a more 
complex situation. In addition to the warm component, the cold component 
itself is structured into a high-density and a low-density contributor.
It is a scaled version of the single-temperature, dual-density model 
(T$_{\rm kin}$ = 10 K; $n(\h2) = 100 \cc$ and 3000 $\cc$) applied to the 
M~31-D~478 cloud complex by Israel et al. (1998). Although a warm component 
must be included, its nature is unclear. Given the large linear beamsize
(1.5 $\times$ 5.5 kpc) in the plane of the galaxy, this warm component may
represent discrete, starforming cloud complexes at some distance from the
center. In all models, $J$=1--0 $\co$ intensities are dominated
by emission from the cold component with contributions of about 75\%,
70\% and 85\% for positions A, B and C respectively. In contrast, the
$J$=3--2 $\co$ intensities are all dominated by emission from the warm
component. For the optically thin $\13co$ transitions the situation
is less clearcut: if we assume an intrinsic isotopic ratio of 100, 
emssion from the cold component contributes about 25\% to the $J$=1--0
emission from positions A and B, whereas this fraction increases to 
about 45\% if we assume an isotopic ratio of 50.

\subsection{CO and C column densities}

In order to relate neutral carbon and carbon monoxide column densities
to that of molecular hydrogen, we have used [C]/[H] gas-phase
abundance ratios estimated from the [O]/[H] abundance. From the data
tabulated by Zaritsky et al. (1994) we determined for
the central beam (position A) 12 + log (O/H) = 9.2, i.e. [O]/[H] = 1.5 
$\times$ 10$^{-3}$. Although high, such an oxygen abundance is normal for 
galaxy centers (Garnett et al. 1997; van Zee et al. 1998). Using results
given by Garnett et al. (1999), notably their Figures 4 and 6, we arrive 
at an estimated carbon abundance [C]/[H] = 2$\pm$1 $\times$ 10$^{-3}$. 
As a significant fraction of all carbon will be tied up in dust particles, 
and not be available in the gas-phase, we adopt a fractional correction 
factor $\delta_{\rm c}$ = 0.33. Neglecting contributions by e.g. $^{13}$CO 
and ionized carbon, we thus find $N_{\rm H}$ = [2$N(H_{2})$ + $N(HI)$] 
$\approx$ 1700 [$N(CO)$ + $N(C)$] with a factor of two uncertainty in 
the numerical factor. Similarly, we find for the off-center positions 
B and C numerical factors of 2300 and 3000. The beam-averaged column 
densities in Table 6 have been obtained by scaling the model cloud 
column densities by the ratio of actual observed CO intensity to predicted 
model CO intensity.

The results of our model calculations are given in Table 6. In the
table we give the predicted [CI] intensity $I_{CI}$, which can be
verified observationally, the calculated beam-averaged column densities
for both CO and C, the $\h2$ column densities derived from these using
the $N_{H}/N_{C}$ ratios and $N(HI)$ values given, as well as the
implied CO to H2 conversion factor $X = N(\h2)/I_{CO}$. The neutral carbon
intensities $I(CI)$ were calculated under the assumption that a significant
fraction (0.6--0.7) of the total atomic carbon column density is ionized
and present in the form of [CII]. Changes in the 
input CO column densities do not strongly affect the resultant C column
density: a substantially higher $N(CO)$, for instance, implies a lower 
$N(C)/N(CO)$ ratio, yielding a relatively unchanged $N(C)$. We have also 
performed the calculations for a ratio of 50. Generally, the ratio of 100 
provides a better fit to the $\13co$ data than the ratio of 50. As the end 
results for the two sets are moreover very similar, we have not included 
the latter in the table. The results of all models are given for position 
A, where we have also measured the [CI] intensity in a 10$''$ beam. However, 
the models apply to measurements in the 21$''$ beam observed or synthesized 
for CO. If atomic carbon is at a minimum in the center, [CI] intensities in
a twice larger beam may be somewhat higher, perhaps by as much as 40$\%$. 
Table 6 shows that model 1 yields a very good fit, whereas models 2 through 
5 are marginally possible and models 6 and 7 are ruled out. Model 1 is not 
unique; various other combinations of somewhat different kinetic temperatures 
for both cold and warm gas and somewhat different densities, yield very
similar results. 

As models 6 and 7 are ruled out for position A and models 2 through 5
yield almost identical final results, we present only models 1 and 3 
for positions B and C. The results for position C have relatively large 
uncertainties due to the weakness of its emission, and the lack of a 
$J$=1--0 $\13co$ measurement. The results are not greatly different 
from those obtained at position A. Column densities decrease, and $X$ 
factors increase somewhat with radius. At $T_{\rm kin}$ = 10 K, the 
$^{3}$P$_{2}$--$^{3}$P$_{1}$ [CI] transition at 809 GHz has negligible 
intensity, but this becomes comparable to the 
$^{3}$P$_{1}$--$^{3}$P$_{0}$ 492 GHz transition at $T_{\rm kin}$ = 30 
K. The presence of the warm component can therefore be verified by 
future observations of the 809 GHz [CI] transition, for which we predict 
an intensity of 15--30\% of the 492 GHz intensity. For the [CII] emission 
we expect intensities of the order of $5 \times 10^{-6}$ erg s$^{-1}$ 
cm$^{-2}$ sr$^{-1}$.

Beam-averaged neutral carbon to carbon monoxide column density ratios are 
$N(C)/N(CO)$ = 0.65$\pm$0.1 and $N(C)/N(CO)$ = 5.5$\pm$1.0 for models 1 
and 3 respectively. The former is close to the typical values 0.2--0.5 
found for M~82, NGC~253 and M~83 (White et al. 1994; Israel et al. 1995; 
Stutzki et al. 1997; Petitpas $\&$ Wilson 1998), but the latter is much 
higher and is only matched by the corresponding ratio of 3--6 found in 
Galactic translucent clouds (Stark $\&$ van Dishoeck 1994). 

\subsection{Molecular hydrogen and the $I(CO)$ to $N(\h2)$ ratio}

Although any explanation of the observed CO intensities requires the 
presence of both cold and smaller amounts of warm molecular gas in 
NGC~7331, the range of admissible parameters is not fully constrained. 
The [CI] intensity observed towards the center of the galaxy, however,
strongly suggests a complex physical environment of the sort represented
by model 1. This model is characterized by cold molecular gas (typical 
temperature $T_{\rm kin}$ = 10 K) present at both high and low volume
densities (typically of order a few hundred and a few thousand per cc 
respectively), in addition to a warmer component (temperature $T_{\rm kin}
\geq 20 K$) of high density. This is probably a simplification: in reality
a range of densities and temperatures is likely to be present. As the
large linear beamsize (1.5 kpc along the major axis, 5.5 kpc along the
minor axis) only provides results averaged over a large radial range,
the spatial distributions of the cold and the warm gas within the beam
may well be different. Both kinetic temperature and mean 
molecular gas density in the centre of NGC~7331 are typically an order of
magnitude below the values found in later-type starburst galaxies such as 
NGC~253 and M~82 (Israel et al. 1995; Wall et al. 1991). 

Our models suggest that a large fraction of the CO emission originates
from cold gas of low column density. A smaller fraction originates in
much denser gas, partly at higher temperatures. The cold gas in the center
of NGC~7331 appears to be similar to that in cloud complexes such as
D~478 in M~31 dark and the Taurus-Auriga complex in the Milky Way; most
likely, it is highly fragmented and filamentary (Israel et al. 1998). 
The warm gas may be heated by the nucleus and by luminous stars in the 
inner spiral arms and the `molecular ring'. The presence of energetic 
photons in the inner part of NGC~7331 is betrayed by H$\alpha$+[NII] 
emission (see Fig. 5 by Smith $\&$ Harvey 1996) and more directly by 
significant UV emission (Wesselius et al. 1982) unlikely to be dominated 
by the spiral arms because of their high dust content (Bianchi et al. 1998).

The evaluation of the models in Tables 5 and 6 assumes a radiation field
$I_{\rm UV} \approx 1$, corresponding to $I_{1000} = 4.5 \times 10^{-8}$
photons s$^{-1}$ cm$^{-2}$ which is consistent with the longer-wavelength 
UV data by Wesselius et al. (1982). Such a low radiation field density 
is also indicated by the strength of the 7.7 and 11.3 $\mu$m dust emission 
features (Smith 1998). The beam-averaged column densities in Table 6 are 
relatively insensitive to changes in the assumed $I_{\rm UV}$, because in 
the cold diffuse gas, most carbon is already in C rather than in CO, 
whereas the much smaller filling factor of the dense gas greatly reduces 
the effect of changes in the $N(C)/N(CO)$ ratio on the beam-averaged
neutral carbon column density. Moreover, we expect only limited variation 
(by a factor of 2--3) in the radiation field density over at least the 
inner 5 kpc because of the smooth distribution of H$\alpha$ emission
as well as the far-infrared emission between 50$\mu$m and 200$\mu$m
(Smith $\&$ Harvey 1996; Alton et al. 1998). The quiescence of the spiral
arms is illustrated by the strong excess of 450 and 850$\mu$m emission from 
cold dust (Bianchi et al. 1998). Thus, the spiral arms contain a relatively
large amount of cold dust especially in comparison with the central region.
We are therefore confident of the derived $N(\h2)$ values in Table 6. 

As C and O abundances in NGC~7331 are 2--5 times higher than those in the 
Solar Neighbourhood, and radiation fields are not particularly intense,
we expect CO in NGC~7331 to be relatively well-shielded, so that the
CO to $\h2$ conversion factor $X$ should be lower than that in the
Solar Neighbourhood, i.e. fewer $\h2$ molecules per unit CO intensity.
Indeed we find values of $X$ lower than the value of $2 \times 10^{20} 
\cm2/\kkms$ assumed for the Milky Way, which can also be compared to the 
relationship between $X$, radiation field intensity and metallicity found 
by Israel (1997). For positions A, B and C we take [O]/[H] abundance ratios 
of 1.5, 1.3 and 1.1 in units of 10$^{-3}$ respectively (Zaritzky et al.
1994). From high-resolution far-infrared surface brightnesses (Smith 
$\&$ Harvey 1996), HI column densities (Begeman (1987) and eqn. (3b) from 
Israel (1997), we predict values $X$ = 0.8, 0.7 and 1.0 in units of $10^{20}$
cm$^{-2}/(\kkms)^{-1}$ for positions A, B and C respectively. These are 
very close to the results from the preferred model 1, and a factor of two 
or more below the results for the other models. Neglect of the radiation 
field term in Israel's (1997) eqn. (3b), i.e. use of his eqn. (4) predicts 
in the same units $X$ = 0.15, 0.25 and 0.4 for positions A, B and C, i.e. 
much lower than any of the model results. We conclude that the low values 
of $X$ in the preferred model 1 are in good agreement with both the high 
abundances in NGC~7331 and the relationship between $X$, radiation field 
intensity and metallicity found by Israel (1997). 

With respect to the value of $X$ derived for position A it should 
be noted that the large linear beamsize includes both the center of 
NGC~7331 and more outlying regions along the minor axis. If the latter 
were to be characterized by an $X$ value closer to that of position B, 
the actual central $X$ value would be significantly lower. For instance, 
if we assign $X = 0.6 \times 10^{20} \cm2/\kkms$ to the outer half of 
the CO emission, the inner half would have $X = 0.2 \times 10^{20} 
\cm2/\kkms$, an order of magnitude less than the Milky Way value, and 
well below what is suggested by the high metallicity. Such a very low 
value would, however, not be unexpected. For the Milky Way centre, 
Sodroski et al. (1995) conclude to an $X$-factor 3 -- 10 times smaller 
than the `standard' Galactic value. The COBE Galactic Centre data 
presented by Bennett et al. (1994) imply lower CO transition ratios 
somewhat similar to those in NGC~7331.

Another way of verifying the derived $\h2$ column densities is provided
by the submmillimeter observations presented by Bianchi et al. (1998). 
For $F_{850\mu}$m = 50 mJy and $T =20\pm3$ K in a $30''\times40''$ beam 
(Bianchi et al. 1998), we derive a beam-averaged $A_{\rm V}$ = 6.6 
(-1.2, +2.2). Furthermore assuming that the dust to gas ratio is 
proportional to metallicity, we modify the Galactic relation between 
total hydrogen column density and visual extinction (Bohlin et al. 1978) 
to $N_{\rm H} = 0.6 \times 10^{21} A_{\rm V}$ cm$^{-2}$. This implies a 
column density $N(\h2) = 2 (-0.4, +0.6) \times 10^{21} \cm2$ (corresponding 
to $X = 0.75 (-0.15, +0.25) \times 10^{20}$). The similarly obtained result
for position B is slightly lower. These results are thus in rather good 
agreement, given the various uncertainties, with $N(\h2) = 1.2-1.6 \times 
10^{21} \cm2$ and $X = 0.4-0.6 \times 10^{20}$ found for positions A and B 
using model 1.

Comparison of the models and the observations allows us to draw some general 
conclusions on the distribution of molecular hydrogen in NGC~7331. In 
model 1, relative amounts of cold/tenuous, cold/dense and warm/dense molecular 
hydrogen gas are 45$\%$, 30$\%$ and 25$\%$ for positions A and B. The 
results for position C seem to indicate a somewhat higher contribution by 
warm molecular gas.  Using the beam-averaged $\h2$ column densities and 
the model $\h2$ volume densities, we find that the {\it average} line of 
sight within the beam contains cold/tenuous $\h2$ over about 2 pc (Model 1, 
positions A and B) to 0.7 pc (Model 1, position C). Both the cold and
the warm dense component have {\it average} line-of-sight extents a factor 
of 50 lower. However, the observed CO temperatures are much lower than
the model excitation temperatures, indicating small beam-filling
factors for the molecular material. Assuming individual lines of
sight within the beam to be either empty, or homogeneously filled 
with molecular gas, we find for those line of sights that do contain
molecular gas extents of about 20 pc (cold tenous gas), 2.5 pc (cold 
dense gas) and 25 pc (warm dense gas); these numbers are indicative
of the maximum source size that can be expected.

Application of model 3 yields somewhat different results. Here, most of the 
molecular gas is in the cold/tenuous form rather than in the warm/dense 
phase: only 17$\%$ warm gas is required at position A, and about 5$\%$ 
at positions B and C.{\it Average} line of sight extents are 3.5 pc for cold
gas at positions A and B, and half that at position C. After correction
for beam filling, we find lines of sight extents of typically 115 pc for 
the cold molecular gas and l5$\%$ or less of that for warm gas. Only at
position C a more uncertain extent of 20 pc is obtained.

The derived line of sight extents are much smaller than the length of
the line of sight traversing the galaxy, which is about four times its 
thickness. Although the latter is not known, this length can be
estimated at well in excess of a kiloparsec. Thus, only a small fraction
of the volume sampled by the beam at each of the analyzed positions
is filled with molecular material. This material is highly clumped or
distributed in filamentary form.

\subsection{Radial distribution of molecular gas}

In the case of highly-inclined ring structures, major-axis position-velocity 
diagrams may give a misleading impression of the actual radial distribution 
of emitting material, because at the tangential points substantially longer 
lines of sight contribute to the emission. To determine the actual 
distribution of CO as a function of radial distance from the centre spectra, 
we have fitted an inclined axisymmetric disk model to the data in the 
velocity-integrated map (Fig. 3) by applying the Richardson-Lucy iterative 
scheme (Lucy 1974). In principle, with {\it a priori} knowledge of the (CO) 
velocity field, this technique can also be used to obtain radial 
distributions with a spatial resolution {\it higher} than that of the 
observing beam (Scoville, Young $\&$ Lucy 1983). Fig. 4 shows the 
fitted radial distribution of the velocity-integrated $J$=2--1 CO emission. 

The fitted profile corresponds to the {\it face-on} radial distribution of 
$I'(CO)_{\rm o}$ = $\int (T_{\rm a}^{*}(CO)dV)_{\rm o}$. The CO luminosity 
starts at $I'(CO)_{\rm o}$ = 4 K $\kms$ in the center, reaches a minimum at
$R \approx$ 1.75 kpc, and the reaches a maximum at $R \approx$ 3.5 kpc after
which it drops smoothly to $I'_{\rm CO}$ = 2 K $\kms$. The ring-to-disk 
intensity contrast ratio is about 0.6. The molecular `ring' is 
clearly discernible, but it does not dominate the CO distribution in the 
galaxy. Both the major-axis CO distribution and the fitted radial CO 
profile are different from those of M~31, where most of the CO is found 
farther out in the spiral arm `ring' at $R$=9 kpc, and very little CO 
occurs at the centre (Dame et al. 1994).

In order to determine the radial distribution of interstellar gas in
NGC~7331, we have converted the CO radial profile to a radial distribution 
of $\h2$ mass densities, using the $J$=1--0/$J$=2--1 CO ratios and $X$ 
values from Tables 4 and 6, and combined these with the radial HI profile 
published by Begeman (1987). Both the radial $\h2$ and HI profiles are 
also shown in Figure 4. In the inner 4.5 kpc, the $\h2$ mass dominates 
that of HI by about 40$\%$. 
Beyond this, HI becomes increasingly dominant. The radial distribution 
of molecular gas reaches its peak (at $R$ =3.1 kpc) well before that of 
HI (at $R \approx$ 10 kpc). Although the radial $J$=2--1 CO profile 
exhibits a relatively low contrast between the ring feature and the 
underlying disk emission, the radially increasing
transitional ratios and $X$-values serve to enhance the contrast in
$\h2$. The central region is not empty, but the $\h2$ mass density inside 
the ring is only 75$\%$ of that in the ring; due to the lack of HI in the 
center, the relative mass density of all hydrogen is even lower with 55$\%$ 
of the ring value. The face-on $\h2$ mass distribution increases from
$\sigma_{\rm \h2}$ = 6 M$_{\sun}$ pc$^{-2}$ to 8 M$_{\sun}$ pc$^{-2}$
at $R$ = 3.1 kpc. The total hydrogen radial mass-density distribution 
increases from a central value $\sigma _{\rm HI+H2}$ = 8 M$_{\sun}$ 
pc$^{-2}$ to $\sigma _{\rm HI+H2}$ = 14 M$_{\sun}$ pc$^{-2}$ at $R$ = 
3.5 kpc and then drops slowly. The gaseous fraction (including helium) of the 
total mass was estimated from the rotation curves given by Rubin et al. 
(1965) and Begeman (1987), assuming a spherical bulge and circular 
velocities. Inside the ring, the gas-to-total mass ratio 
$M_{\rm gas}$/$M_{\rm dyn}$ is about 1$\%$. In the ring, it rises to 
1.5$\%$, and then slowly climbs to 3$\%$. From Begeman's (1987) data,
neglecting $\h2$, we find in comparison a {\it global} ratio 
$M_{\rm gas}$/$M_{\rm dyn}$ of 3.2$\%$. Even with the 
dominant contribution by H$_{2}$, the gas in the inner part of NGC~7331 
is only a minute fraction of the total mass. 

\begin{figure}
\hspace{2.0cm}
\resizebox{12cm}{!}{\rotatebox{270}{\includegraphics*{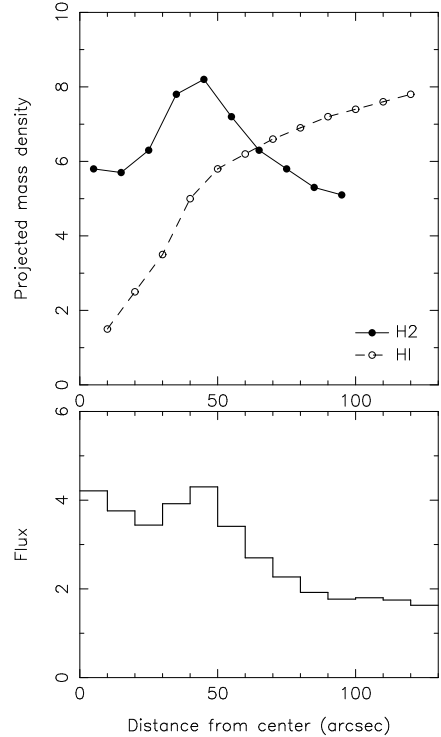}}}
\caption[]{Deprojected radial profiles. Bottom: Face-on radial distribution 
of CO emission obtained with the Richardson-Lucy scheme (see section 4.4). 
Vertical axis is $\int T_{\rm mb}$d$V$ as would be observed perpendicular 
to the galaxy plane. Top: Face-on mass-densities of $\h2$ and HI, in units
of M$_{\odot}$ pc$^{-2}$. HI data were taken from Begeman (1987).
}
\end{figure}

Finally, it is of interest to compare the radial gas distributions to
the radial far-infrared profiles representing interstellar dust.
The 100$\mu$m far-infrared profile (Smith $\&$ Harvey 1996) is rather
flat in the center, reaching a very minor maximum at about 2 kpc,
after which a steep decline sets in. The 450$\mu$ and 850$\mu$m
profiles show radially increasing intensities that reach a peak at
about 3 kpc, the 850$\mu$m profile peaking a little farther out than
the 450 $\mu$m profile (Bianchi et al. 1998). The $\h2$ mass-density 
peaks at $R$ = 3.1 kpc, just about coincident with the 850$\mu$m 
maximum. At the location of the molecular `ring' peak, 100$\mu$m 
intensities are about 80$\%$ of those in the center. However, the total gas 
mass-density peaks slightly farther out at $R$ = 3.8 kpc. Thus, emission 
from warm dust peaks well inside the molecular ring, and both molecular 
gas and cold dust peak inside the radius of highest gas mass-density. 
Beyond the ring peak, the decline of the 450$\mu$ and 850$\mu$m emission
more or less follows that of the $\h2$ mass density profile. 
We conclude that the interstellar dust is 
hottest in the central region, where gas mass-densities are lowest. 
The mean dust temperature defined by the far-infrared ratios smoothly 
decreases from the center reaching a shallow minimum at about $R$ = 3 
kpc, i.e. at the {\it H$_{2}$ peak}, beyond which it appears to increase 
slightly.

\subsection{Origin of bulge molecular gas and the ring}

Various mechanisms for the occurence of ring morphologies have been
suggested in the literature. Interaction between magnetic fields and 
the gas distribution were proposed and tested by Battaner et al. (1988). 
Inner Lindblad resonances may be responsible (Kormendy $\&$ Norman 1979), 
and a ringlike feature may also result from evacuation of gas from the 
central regions by stellar winds (cf. Faber $\&$ Gallagher 1976; 
Soifer et al. 1986, Mu\~noz-Tu\~non $\&$ Beckman 1988). Finally, as
Young $\&$ Scoville (1982) suggested for NGC~7331, a ringlike appearance
may also be caused by the nuclear bulge having used up the originally
present molecular gas in the center. 

In NGC~7331, the solid-body rotation curve rises rapidly out to $R$ = 3.5 
kpc after which it flattens and reaches a broad maximum at $R$ = 6 kpc 
(see also von Linden et al. 1996). The ring is thus located just at the 
radius where rigid rotation is lost, but well within the radius of maximum 
rotational velocity ($R_{\rm ring}$ = 0.6 $R_{\rm Vmax}$). This is unlike 
M~31, where the molecular ring is found at a radius {\it twice} that of peak 
rotational velocity (Brinks $\&$ Burton 1984; Dame et al. 1993). Both the 
molecular ring and the boundary of the solid-body rotation region are also
just at the radius at which the light of the disk becomes dominant over 
that of the bulge (cf. Begeman 1987). Although the radius of the 
mostly nonthermal radio continuum ring radius is more difficult to determine 
because of its inhomogeneous structure, it appears to coincide more or
less with the molecular ring, also well inside the radius of maximum 
rotational velocity (Cowan et al. 1994).

On the same reasoning as used by Young $\&$ Scoville (1982), we may rule out 
the presence of an inner Lindblad resonance in NGC~7331 as an explanation for 
the observed molecular ring structure, because an ILR can only occur well 
outside the region of solid-body rotation (e.g. Kormendy $\&$ Norman 1979).
Our observations clearly show that there is no pronounced CO hole in the 
centre of NGC~7331, although {\it the derived distributions of both $\h2$
and total interstellar gas do show a significant central depression}. 
von Linden et al. (1996) have suggested that the ringlike distribution in
NGC~7331 is caused by the dynamical action of a weak central bar. However, 
their simulations yield both a ring more massive than observed, and center 
more devoid of gas than observed, casting doubts on the proposed
central bar.

In the case of M~31, Soifer et al. (1986) suggest that the amount of 
interstellar matter observed in the center of M~31 could have accumulated 
from late-type stellar mass loss in the bulge, and is kept low by 
continuous gas removal by supernova explosions and star formation. 
Could this also be the case in NGC~7331? It has been suggested by Prada
et al. (1996) that the bulge of NGC~7331 is counter-rotating. Since the 
bulge gas is rotating in the normal sense, this would seem to preclude
a stellar origin for the gas. However, spectroscopy by Mediavilla et 
al. (1997) and Bottema (1999) does not confirm the suggested 
counter-rotation. The total 
mass of the interstellar gas inside $R$ = 2 kpc is 1.6 $\times$ 10$^{8}$ 
M$_{\sun}$. According to the reasoning outlined by Soifer et al. 
(1986), stellar mass loss in the bulge would accumulate this amount in 
4 $\times$ 10$^{8}$ years. Removal of the same amount of material from 
the bulge requires the energy output of $6/n \times 10^{5}$ Type I 
supernovae, $n$ being the fraction of energy available for the acceleration 
of interstellar material. For a Type I SN rate of 4 $\times$ 10$^{-13}$ 
$L_{\rm B}$ yr$^{-1}$ (Lang, 1992) the timescale for removal is 6.5/n 
$\times$ 10$^{7}$ years. Only if the fraction of supernova energy actually
available for removal exceeds 17$\%$, will the interstellar gas be evacuated 
from the bulge faster than bulge stars can manufacture it. More generally, 
with the assumptions from Soifer et al. (1986), the ratio of evacuation to 
deposition timescales is $t_{e}/t_{d}$ = $3.75 \times 10^{-6} 
v_{\rm c}^{2}/n$. The tabulation of NGC~7331 rotation velocities by 
Begeman (1987) then suggests relatively efficient evacuation in the 
inner 1 kpc ($t_{e}/t_{d} = 0.01/n - 0.05/n)$, and much less efficient 
evacuation at the edge of the bulge ($R$ = 4 kpc; $t_{e}/t_{d} = 0.24/n$). 
The radial decrease of the ratio of far-infrared emission to $\h2$ 
mass-density found above may be related to this finding. It thus appears 
that the relatively small amounts of interstellar gas in the bulge of 
NGC~7331 ($M_{\rm gas}/M_{\rm dyn} \approx 0.01$) also may 
well be the result of mass loss from the bulge stars themselves, rather 
than the result from a net inflow of molecular material from greater 
radii. 

\section{Conclusions}

\begin{enumerate}

\item Analysis of the $J$=2--1 $^{12}$CO distribution and kinematics
shows the presence of enhanced molecular emission in a ringlike zone in 
NGC~7331, peaking at a radial distance of 3.5 kpc with a width of about 
2 kpc. At $R$ = 3.5 kpc, the velocity-integrated CO intensity of the ring 
itself is about 0.6 times that of the underlying more smoothly distributed 
CO emission that fills the entire bulge of NGC~7331.

\item The velocity-integrated CO intensities in the center of NGC~7331
decrease strongly with increasing rotational level. The intensities in
the $J$=1--0, $J$=2--1, $J$=3--2 transitions are in the ratio of
1.0 : 0.55 : 0.35 respectively. The observed $^{12}$CO/$^{13}$CO 
isotopic ratios are 6.7 and 5.6 in the $J$=1--0 and $J$=2--1 transitions
respectively. Positions at larger radial distances have similar ratios,
albeit with somewhat stronger $J$=3--2 CO emission, and weaker
$^{13}$CO emission. Weak [CI] emission was detected from the center.

\item Modelling of the observed line ratios suggest a multi-component
molecular medium. Gas with a kinetic temperature of about 10 K appears to
be present at both low and high densities. At high densities, a warmer 
component with a kinetic temperature of 20 K or more is also present
within the observing beams. The gas is probably distributed in a 
clumpy and filamentary form.

\item Assuming a [C]/[H] abundance ratio of the order of 1--2 $\times 
10^{-3}$, the mean CO-to-H$_{2}$ conversion factor is $X_{\rm N7331}$ = 
4 $\times$ 10$^{19}$ cm$^{-2}$ in the bulge region, and double that
value in the ring and beyond. These values are well below those found in 
the Solar Neighbourhood, but they are consistent with the high metallicity
of NGC~7331 and with submillimeter dust observations.

\item In the bulge, interstellar gas (HI + H$_{2}$ + He) mass densities, 
projected onto the plane of the galaxy, are of the order of 11 M$_{\odot}$ 
cm$^{-2}$. In the ring itself, now properly placed at $R$ = 3.1 kpc, 
the gas mass density is almost twice as high. Within 
the ring, the interstellar gas mass is dominated by the molecular hydrogen 
contribution. Gas to total (dynamical) mass ratios are about 1 $\%$ in the 
center and about 1.5 $\%$ in the ring.

\item The molecular ring coincides more or less with the mostly nonthermal 
radio continuum ring and the 850 $\mu$m ring representing emission from
cold dust. Emission from warmer dust in the 100$\mu$m wavelength range 
peaks well inside the molecular ring; dust temperatures appear to be 
decreasing with radius reaching a mininmum in the ring. The radial 
distribution of HI reaches it maximum well beyond the molecular ring.

\item The molecular ring is well inside the radius of peak rotational
velocity. Its maximum is just at the edge of the region of solid-body 
rotation, and just at the radius where disk light becomes dominant
over bulge light. The ring is not associated with an inner Lindblad 
resonance. The molecular gas inside the ring may have originated
from mass loss by late type stars in the bulge. If this is the case, the
ring is probably the result of wind-driven gas removal from the center.

\end{enumerate}
\acknowledgements

We are indebted to Ewine van Dishoeck and David Jansen for providing us 
with their detailed radiative transfer models. We also thank the JCMT 
personnel, in particular Remo Tilanus, for their support and help in 
obtaining the observations discussed in this paper, and Jeroen
Stil for considerable help in producing Figure 2. The IRAM observations 
were kindly obtained for us by Gabriel Paubert in service mode. 


\begin{thebibliography}{}
%
\bibitem{} Aalto S., Black J.H., Johansson L.E.B., Booth R.S., 1991, \aua 
	249, 323
\bibitem{} Allen R.J., Lequeux J., 1993, \apjl 410, L15
\bibitem{} Allen R.J., Le Bourlot J., Lequeux J., Pineau des For\^ets
	G., Roueff E., 1995, \apj 444, 157
\bibitem{} Alton P.B., Trewhella M, Davies, J.I. et al., 1998, \aua 335, 807
\bibitem{} Arp H.C., Kormendy J., 1972, \apjl 178, L101
\bibitem{} Battaner E., Florido E., Sanchez-Saavedra M.L., 1988, \apj 331, 116
\bibitem{} Begeman K., 1987, Ph.D. Thesis, University of Groningen (NL)
\bibitem{} Bennett C.L., Fixsen D.J., Hinshaw G., et al. 1994, \apj 434, 587
\bibitem{} Bianchi S., Alton P.B., Davies J.I., Trewhella M. 1998, \mnras 298, 
	L49
\bibitem{} Bohlin R.C., Savage B.D., Drake J.F., 1978, \apj 224, 132
\bibitem{} Bosma A.: 1978, Ph.D. thesis, University of Groningen (NL)
\bibitem{} Bottema R.: 1999, \aua 348, 77
\bibitem{} Braine, J., Combes F., Casoli F., et al., 1993, \apjs 97, 887
\bibitem{} Brinks E., Burton W.B., 1984, \aua 141, 195
\bibitem{} Cowan J.J., Romanishin W., Branch D., 1994, \apjl 436, L139
\bibitem{} Dame T.M., Koper E., Israel F.P., Thaddeus P., 1994, \apj 418, 730
\bibitem{} Dressel L.L., Condon J.J. 1976, \apjs 31, 187
\bibitem{} Elfhag T., Booth R.S., H\"oglund B., Johansson L.E.B., Sandquist Aa.,
	1996, \auas 115, 439
\bibitem{} Faber S.M., Gallagher J.S., 1976, \apj 204, 365
\bibitem{} Garnett D.R., Shields G.A., Skillman E.D., Sagan S.P., Dufour R.J., 
	1997, \apj 489, 63
\bibitem{} Garnett D.R., Shields G.A., Peimbert M., et al. 1999 \apj 513, 168
\bibitem{} Israel F.P., 1997, \aua 328, 471
\bibitem{} Israel F.P., White G.J., Baas F., 1995, \aua 302, 343
\bibitem{} Israel F.P., van der Werf P.P., 1996, in: {\it Cold Gas at High
	Redshift}, ed. M.N. Bremer, P.P. van der Werf, H.J.A. R\"ottgering,
	C.L. Carilli (Dordrecht: Kluwer), p. 429
\bibitem{} Israel F.P., Tilanus R.P.J., Baas F., 1998, \aua 339, 398
\bibitem{} Jansen D.J., 1995, Ph.D. thesis, University of Leiden (NL)
\bibitem{} Jansen D.J., van Dishoeck E.F., Black J.H., 1994, \aua, 282, 605
\bibitem{} Koper E. 1993, Ph.D. thesis, University of Leiden (NL)
\bibitem{} Kormendy J., Norman C.A., 1979, \apj 233, 539
\bibitem{} Lang K.R., 1992, {\it Astrophysical Data}, (New York: Springer), 
	p. 703
\bibitem{} Loinard L., Allen R.J., Lequeux J., 1995, \aua 301, 68
\bibitem{} Loinard L., Allen R.J., 1998, \apjl 499, 277
\bibitem{} Lucy L.B., 1974, \apj 79, 745
\bibitem{} Mediavilla E., Arribas S., Garc\'ia-Lorenzo B., del Burgo C.,
	1997, \apj 488, 682
\bibitem{} Mu\~noz-Tu\~non C., Beckman J.E., 1988, {\rm Ap$\&$SS}, 147, 173
\bibitem{} Petitpas G.R., Wilson C.D., 1998, \apj 503, 219
\bibitem{} Prada F., Guti\'errez C.M., Peletier R.F., McKeith C.D., 1996,
	\apjl 463, L9
\bibitem{} Rice W., Lonsdale C.J., Soifer B.T., et al., 1988, \apjs 68, 91
\bibitem{} Rubin V.C., Burbidge E.M., Burbidge G.R., Crampin D.J., 1965,
	\apj 141, 759
\bibitem{} Sandage A., Tammann G.A., 1987, {\it A Revised Sghapley-Ames
	Catalog of Bright Galaxies}, second edition, Cargegie Institution of
	Washington Publication 635 (Washington, D.C.: Carnegie Institution of
	Washington).
\bibitem{} Sandage A., Bedke J., 1988, {\it Atlas of Galaxies}, NASA-SP 496
	(Washington, D.C.: NASA).
\bibitem{} Scoville N.Z., Solomon P.M., 1975, \apjl 199, L105
\bibitem{} Scoville N.Z., Young J.S., Lucy L.B., 1983, \apj 270, 443
\bibitem{} Smith B.J., 1998, \apj 500, 181
\bibitem{} Smith B.J., Harvey P.M., 1996, \apj 468, 139
\bibitem{} Sodroski T.J., Odegard N., Dwek E., et al., 1995, \apj 452, 262
\bibitem{} Soifer B.T., Rice W.L., Mould J.R., et al., 1986, \apj 304, 651
\bibitem{} Stark A.A. 1979, Ph.D. thesis, Princeton University (USA)
\bibitem{} Stark R., van Dishoeck E.F., 1994, \aua 286, 443
\bibitem{} Stockdale C.J., Romanishin W., Cowan J.J, 1998 \apj 508, 33
\bibitem{} Stutzki J., Graf U.U., Honingh C.E., et al. 1997, \apjl 477, 33
\bibitem{} Telesco C.M., Gatley I., Steward J.M., 1982, \apj 263, L13
\bibitem{} Tosaki T., Shioya Y., 1997, \apj 484, 664
\bibitem{} Tully R.B., 1988, {\it Nearby Galaxies Catalog}, (Cambridge:
	Cambridge University Press)
\bibitem{} van Dishoeck E.F., Black J.H., 1988, ApJ 334, 771
\bibitem{} van Zee L., Salzer J.J., Haynes M.P., O'Donoghue A.A., Balonek T.J.,
	1998 \aj 116, 280
\bibitem{} von Linden S., Reuter H.-P, Heidt J., Wielebinski R., Pohl M.,
	1996, \aua 315, 52
\bibitem{} Wall W.F., Jaffe D.T., Israel F.P., Bash F.N., 1991, \apj 380, 384
\bibitem{} Wesselius P.R., van Duinen R.J., de Jonge A.W., et al., 1982,
	\auas 49, 427
\bibitem{} White G.J., Ellison B., Claude S., Dent W.R.F., Matheson D.N.,
	1994, \aua 284, L23
\bibitem{} Young J.S., Scoville, N.Z., 1982, \apjl 260, L41
\bibitem{} Young J.S., Xie S., Tacconi L., et al. 1995, \apjs 98, 219
\bibitem{} Zaritsky D., Kennicutt R.C., Huchra J.P., 1994, \apj 420, 87
%
\end{thebibliography}
\end{document}